\documentclass[final,authoryear,3p,times]{elsarticle}

\usepackage{txfonts}

\usepackage{amssymb}
\usepackage{epsfig}
\usepackage{import}
\usepackage{subfig}
\usepackage{graphicx}
\usepackage{tabularx}
\usepackage{placeins}
\newcolumntype{M}[1]{>{\centering\arraybackslash}m{#1}}

\journal{Journal of the Mechanics and Physics of Solids}

\begin{document}

\begin{frontmatter}

\title{Finite size effects on crack front pinning at heterogeneous planar interfaces: experimental, 
finite elements and perturbation approaches}

\author[a]{S. Patinet}
\author[b]{L. Alzate}
\author[b]{E. Barthel}
\author[b]{D. Dalmas}
\author[a]{D. Vandembroucq}
\author[c]{V. Lazarus}

\address[a]{Laboratoire PMMH, UMR 7636 CNRS/ESPCI/P6/P7, 10 rue Vauquelin, 75231 Paris cedex 05, France}
\address[b]{Surface du verre et interfaces, Unit\'{e} Mixte CNRS / Saint-Gobain UMR 125 (SVI), 39 quai Lucien Lefranc, 93300 Aubervilliers, France}
\address[c]{UPMC Univ Paris 6, Univ Paris-Sud, CNRS, UMR 7608, Lab FAST, Bat 502, Campus Univ, F-91405 Orsay, France}

\begin{abstract}
Understanding the role played by the microstructure of materials on their macroscopic failure properties is an important challenge in solid mechanics. Indeed, when a crack propagates at a heterogeneous brittle interface, the front is trapped by tougher regions and deforms. This pinning induces non-linearities in the crack propagation problem, even within Linear Elastic Fracture Mechanics theory, and modifies the overall failure properties of the material. For example crack front pinning by tougher places could increase the fracture resistance of multilayer structures, with interesting technological applications. Analytical perturbation approaches, based on Bueckner-Rice elastic line models, focus on the crack front perturbations, hence allow for a description of these phenomena. Here, they are applied to experiments investigating the propagation of a purely interfacial crack in a simple toughness pattern: a single defect strip surrounded by homogeneous interface. We show that by taking into account the finite size of the body, quantitative agreement with experimental and finite elements results is achieved. In particular this method allows to predict the toughness contrast, i.e. the toughness difference between the single defect strip and its homogeneous surrounding medium. This opens the way to a more accurate use of the perturbation method to study more disordered heterogeneous materials, where the finite elements method is less adequate. From our results, we also propose a simple method to determine the adhesion energy of tough interfaces by measuring the crack front deformation induced by known interface patterns.
\end{abstract}

\begin{keyword}
Interfacial brittle fracture \sep Toughening \sep Crack pinning
\sep Finite element method \sep Perturbation approach
\end{keyword}

\end{frontmatter}

\section{Introduction}

Predicting the threshold for crack propagation is a key issue in material
science: it determines the design quality of structures and their durability
for a wide range of systems ranging from bulk materials to thin films. 
In particular it was shown that one of the most efficient mechanisms to increase
effective toughness is crack pinning by material heterogeneities~\citep{Bower1991}. 
For example the macroscopic effective toughness of a material can be increased very 
significantly by the dispersion of hard particles in the matrix \citep{Mower1995}. 
Another system that takes advantage of crack pinning consists of patterned interfaces presenting 
heterogeneous toughness landscapes. Their technological interest lies in the development 
of multi-layer materials with high mechanical stability. A promising method to develop new materials consists 
in designing optimal heterogeneous interfaces with high toughness while maintaining their 
functional features. 

Heterogeneities affect the effective toughness by interfering with crack propagation, 
deflecting crack fronts and crack surfaces. However, it is difficult to predict the 
effect of heterogeneities quantitatively and experimental investigation of this toughening 
mechanism raises several difficulties. For example it is not easy to create well controlled
heterogeneity distributions and to observe crack propagation \emph{in situ}.
Also, due to crack deflection, the problem is often three-dimensional and
theoretical or numerical developments become quite complex.

In the literature, most of the experimental investigations on crack pinning in
heterogeneous materials were statistical approaches and studied post-mortem
fracture surfaces \citep{Bouchaud1997,Santucci07,Ponson2007,Dalmas08,Bonamy09}.
Experiments with direct visualization of the crack front \emph{in situ} during
propagation are very unfrequent. For example, with an original experimental
setup, \cite{Schmittbuhl1997} were able to observe the crack front morphology
during propagation along a disordered heterogeneous interface. This setup has
been widely used to obtain statistical information on crack front roughness
\citep{Delaplace1999, Santucci2010} and stochastic dynamics of propagation
\citep{Maloy2001, TalTouSan11}. However quantitative predictions seem out of
reach, mainly because of the presence of shear on the crack front during
loading and the ill-controlled nature of the heterogeneities. Recently,
\cite{Chopin2011} proposed a new experimental approach with better controlled
heterogeneities, but the setup also suffers from mode mixity. In the
experiments of \cite{Mower1995}, the front is trapped  by second-phase
particles introduced in a brittle epoxy. The front shape becomes complicated
since it can not penetrate in the particles, and these experiments are
difficult to model analytically.

Here, we report on a study of crack propagation measured along a well
controlled patterned interface in a classical Double Cantilever Beam
(DCB) geometry. The crack propagates at the weakest interface in a
stack of thin films deposited on glass, allowing direct
visualization of the crack front. This weakest interface is not
homogeneous however: a defect strip with a different interfacial
toughness lies at the center of the sample. Several values of
toughness contrast have been investigated and we have monitored the
interaction of the interface crack with these defects. The interface
cracks are either trapped or attracted by the defect depending on
the sign of the toughness contrast between the defect strip and its 
homogeneous surrounding medium. This setup \citep{Barthel2005,Dalmas2009} 
has several advantages: 1) the crack is loaded in pure tensile mode (mode I); 2)
the crack propagation is purely interfacial (without deflexion out
of the plane of the interface); 3) the toughness contrast can be
tuned to keep the deformations of the crack front small. 

In the framework of Linear Elastic Fracture Mechanics, propagation 
criteria are based on the 
comparison of the Energy Release Rate (ERR) and adhesion energy
\citep{Griffith21} \--- or equivalently of mode I stress intensity
factor (SIF) and toughness~\citep{Irwin58}. The ERR is derived from
an elastic analysis, and depends on loading, elastic response of the
solid and geometry of both body \emph{and} crack. Perturbation
methods were initiated by \cite{Rice1985}, based
on \cite{Bueckner1987} weight-function theory, and specifically used by
\cite{GaoRice89} to study the morphology of cracks trapped in
heterogeneous interfaces. They provide analytical predictions for
the variation of the ERR when the crack front is slightly perturbed
and effectively describe the crack front as an elastic line.
However, the first order formula proposed by \cite{Rice1985} for a
half-plane crack in an infinite media does not take into account any
finite size effect. In the case of the present measurements, they
may provide erroneous predictions for large enough crack front
perturbations (deformation or wavelength) in comparison with sample
sizes. However several improvements of this formula have been
proposed to take into account the finite size of the crack
\citep{GaoRice87b, Gao88, Leblond1996, Lazarus2002, Lazarus2011},
the finite thickness of the body \citep{Legrand2011} and even the
interaction between two cracks \citep{Pindra2010}.

Our aim here is to assess the use of analytical perturbation
theories to model crack propagation in heterogeneous interfaces. To
evaluate the merits of these models, we need a reference solution
which would provide an \emph{exact} modeling of the data. In the
case of the mildly distorted crack fronts proposed here, Finite
Element (FE) simulations can be used to determine the \emph{local}
values of the ERR. In addition, with FE
simulations, we can accurately take into account the exact geometry
of the sample, including \emph{the shape of the crack front} which is
determined experimentally.

The experimental method is presented first with some details concerning the 
synthesis of the samples and the cleavage tests (section \ref{se:exp}). From 
the measured crack front deformations 
and loadings, the ERR pattern is then determined by numerical and 
theoretical methods. The ERR along the crack front is quantitatively
calculated through detailed three-dimensional FE analysis
(section~\ref{se:FE}). Analytical perturbation methods are
presented in section~\ref{se:th} and we show how the local
ERR contrast can be inferred. Then FE and analytical methods
are applied to the different samples and
their results compared in section~\ref{se:comp}. In particular, it is shown that the
ERR contrasts calculated by the two approaches are in good
agreement if the finite size of the plate is duly taken into account
in the analytical perturbation models~\citep{Legrand2011}, in
contrast to \cite{Rice1985}'s model, which is conventionally used
for this type of geometry although it can apply only to infinite
half-spaces. We finally discuss the crack propagation criterion and show how 
these results can be used to give quantitative estimates of local adhesion 
energy values for textured interfaces in section~\ref{se:properties}.

\section{Experiments}
\label{se:exp}

\subsection{Synthesis of samples}

\begin{figure}[h!]
\begin{center}
\textbf{(a)}
\includegraphics[width=0.5\textwidth]{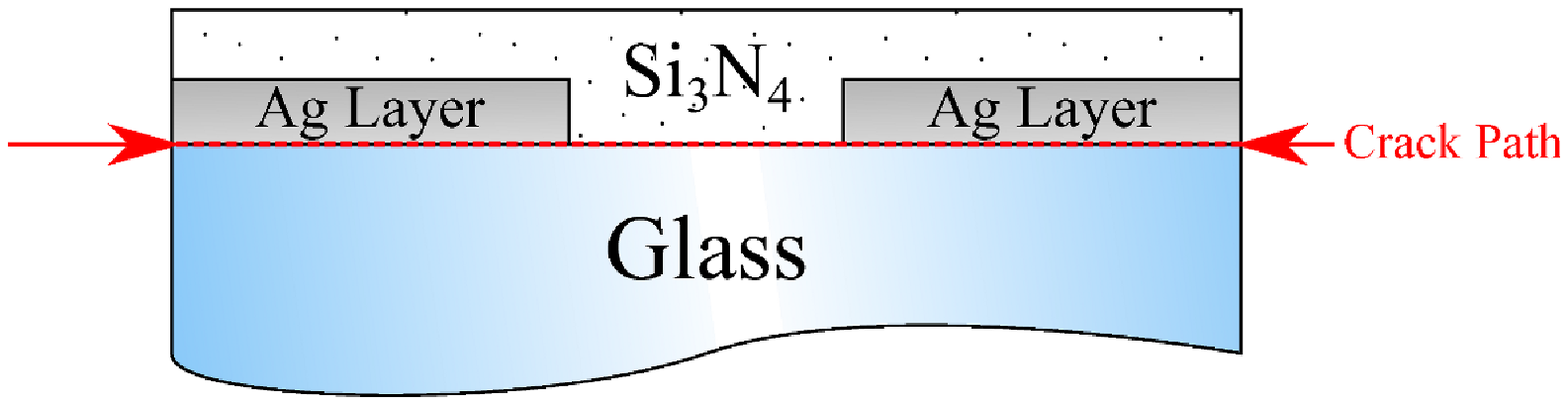}\\
\end{center}  
\textbf{(b)}
\includegraphics[width=0.30\textwidth]{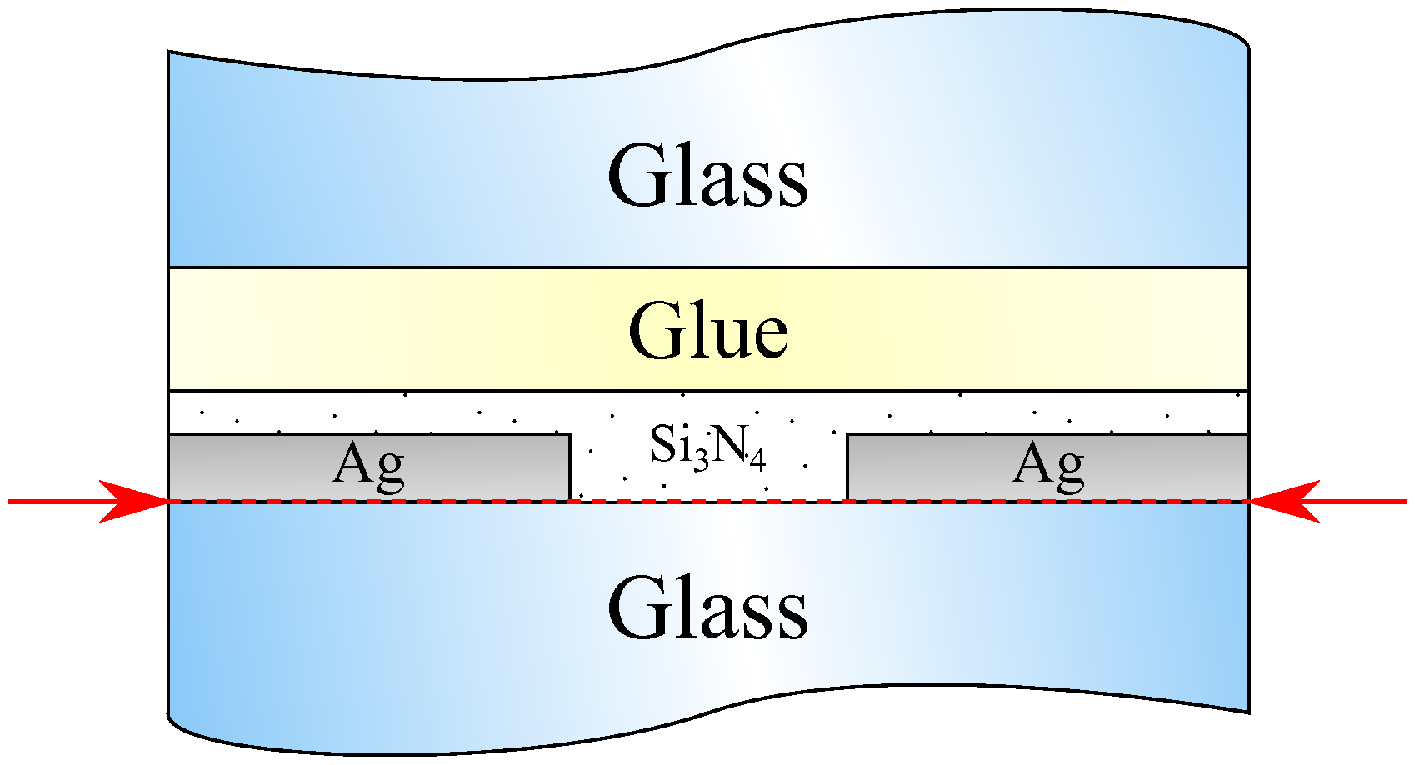}
\textbf{(c)}
\includegraphics[width=0.30\textwidth]{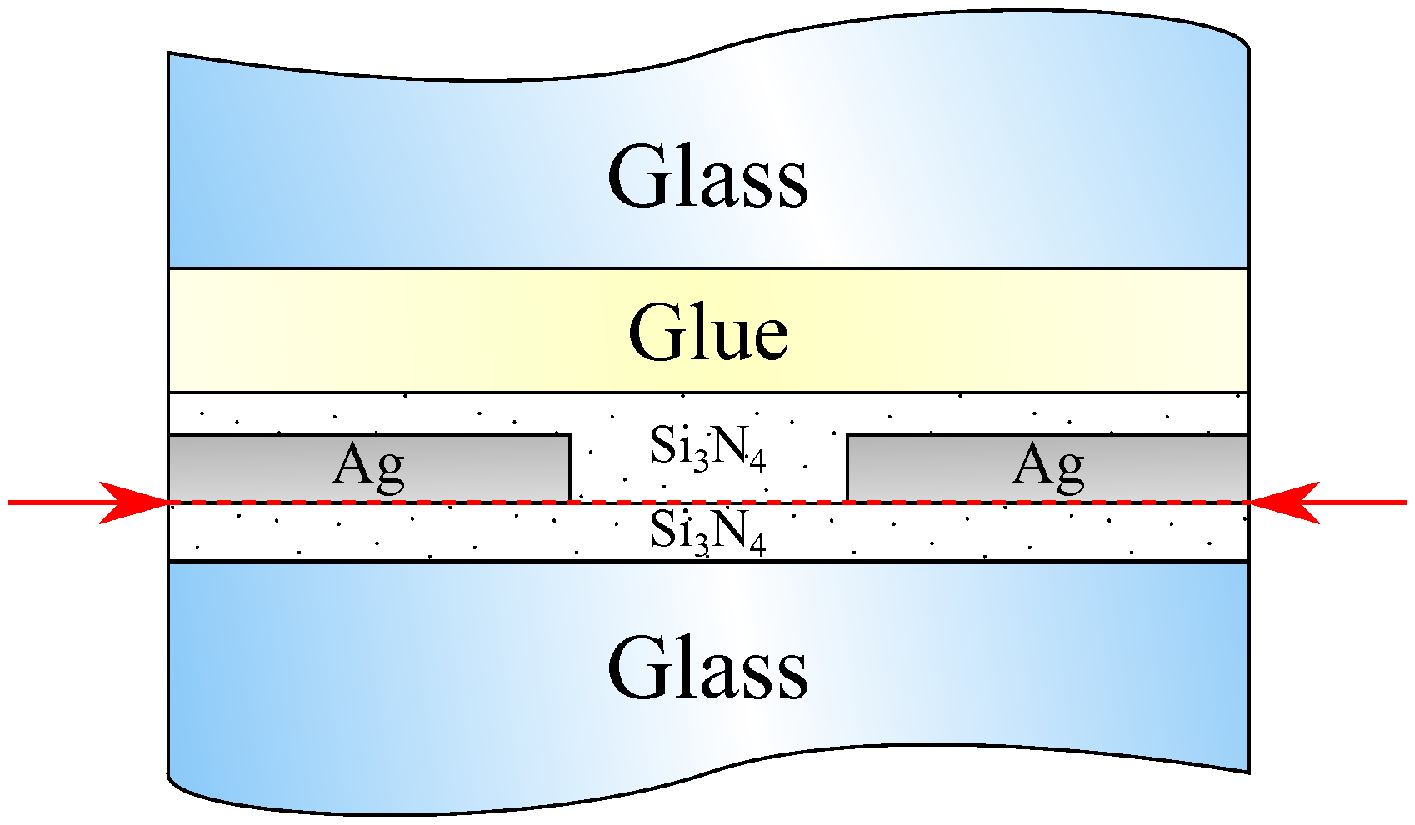}
\textbf{(d)}
\includegraphics[width=0.30\textwidth]{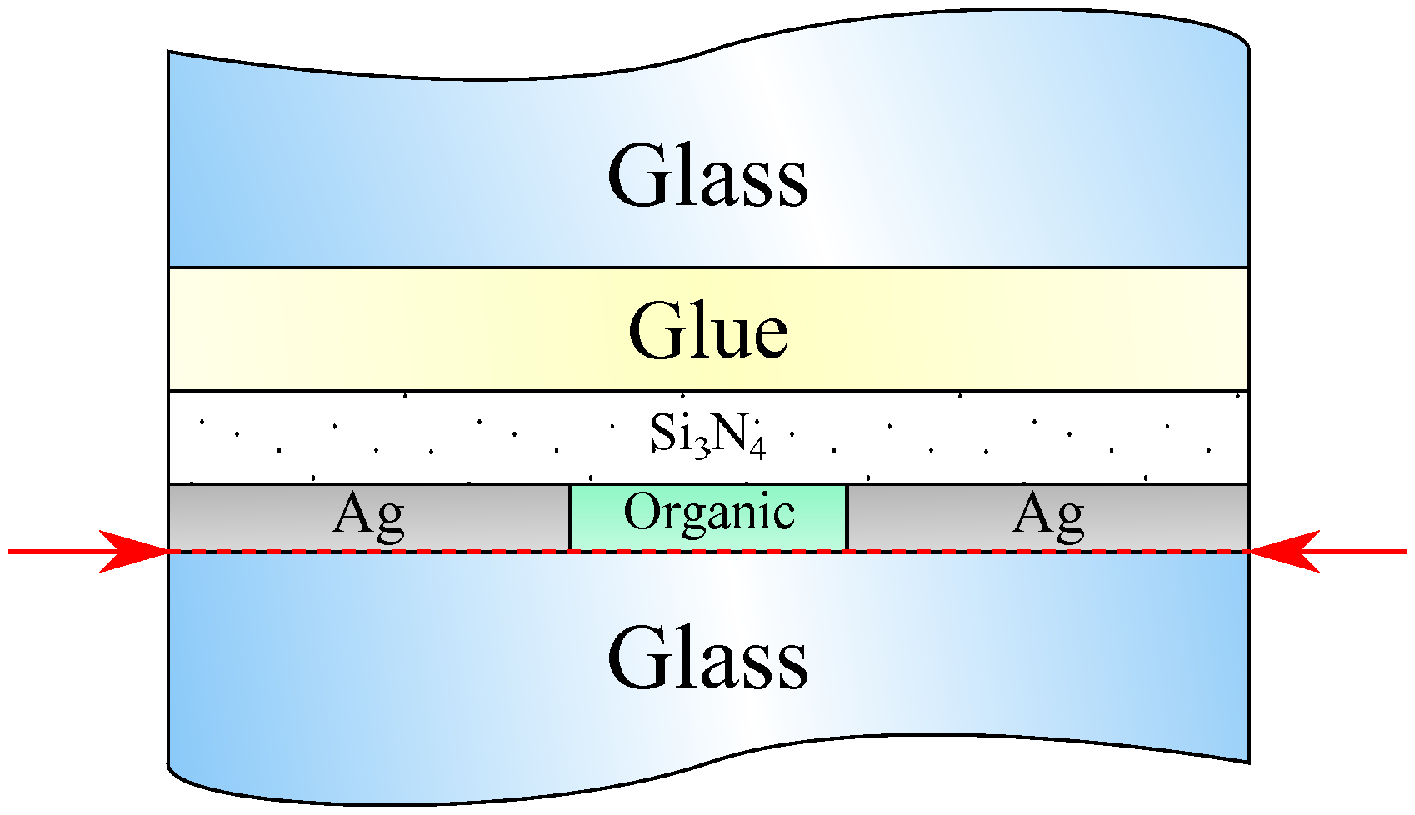}
\caption{(a) Schematics of a patterned multilayer deposited on glass
exemplified for the HC-b30 sample. (b), (c) and (d): the multi-layers are
reinforced by a glass backing prior to cleavage test for the HC-b30, MC-b40 (MC-b50) 
and LC-b40 samples respectively (see text for details). The plane of fracture is shown in red.}
\label{fig:multilayer}
\end{figure}

To control and visualize the propagation of a purely interfacial
crack along a patterned interface we start from thin film
multi-layers deposited by magnetron sputtering~\citep{Barthel2005}
on rectangular glass substrates (thickness $h=700$~$\mu$m, width
$b$, length $L$). Schematically, a typical sample is
glass/silver/top layer (see Fig.~\ref{fig:multilayer}).

Patterning of the interface is achieved by inserting a mask close to
the sample during deposition of the silver layer. The mask is
rectangular (width $d$) and extends over about half the sample
length. The other layers comprising the sample are deposited without
mask and are homogeneous. The area covered by the mask induces a
zone with different adhesion, i.e. different toughness , which we call the defect strip.

Three types of interfaces are studied, with respectively low
toughness contrast (LC), medium contrast (MC) and high contrast
(HC). In LC-b40, the defect strip is an organic layer which is
directly deposited on glass, inducing lower adhesion. For MC-b40 and
MC-b50, the silver layer is deposited on an Si$_3$N$_4$ sublayer.
The toughness contrast originates from the difference between
Ag/Si$_3$N$_4$ and Si$_3$N$_4$/Si$_3$N$_4$ adhesion. In HC-b40 the
silver layer was deposited directly on the glass substrate,
increasing the toughness contrast due to lower adhesion on glass.
The sample nature and dimensions are summarized in
Tab.~\ref{ta:stack}, ranked from the lowest to the highest adhesion.

\begin{table}[h!]
    \centering
    {\small
        \begin{tabular}{|c|c|c|c|c|c|c|c|}
        \hline
         Sample & Stack & $b$ (mm) & $L$ (mm) & $d$ (mm) & $H$ (mm) &  Homog. Interf. & Defect strip\\ \hline
         LC-b40  & Gl./\textbf{Org/Ag}/Si$_3$N$_4$ & 38 & 64 & 2.82 & 3 & Glass/Ag & Glass/Organic\\ \hline
         MC-b40  & Gl./Si$_3$N$_4$/\textbf{Ag}/Si$_3$N$_4$ & 40 & 65 & 3.2 & 3 & Si$_3$N$_4$/Ag & Si$_3$N$_4$/Si$_3$N$_4$ \\ \hline
         MC-b50  & Gl./Si$_3$N$_4$/\textbf{Ag}/Si$_3$N$_4$ & 49 & 82.5 & 3.87 & 2.5 & Si$_3$N$_4$/Ag & Si$_3$N$_4$/Si$_3$N$_4$ \\ \hline
         HC-b30  & Gl./\textbf{Ag}/Si$_3$N$_4$ & 29.5 & 36.5 & 3.2 & 1.5 & Glass/Ag & Glass/Si$_3$N$_4$  \\ \hline
      \end{tabular}
      }
        \caption{Nature and dimensions of the patterned samples. For each stack the patterned layer 
        is shown in bold. Underlayer thicknesses vary between 10 and 50~nm.}
\label{ta:stack}
\end{table}

\subsection{Cleavage test}

\begin{figure}[h!]
\begin{minipage}{1.\textwidth}
  $\vcenter{\hbox{\includegraphics[width=0.64\textwidth]{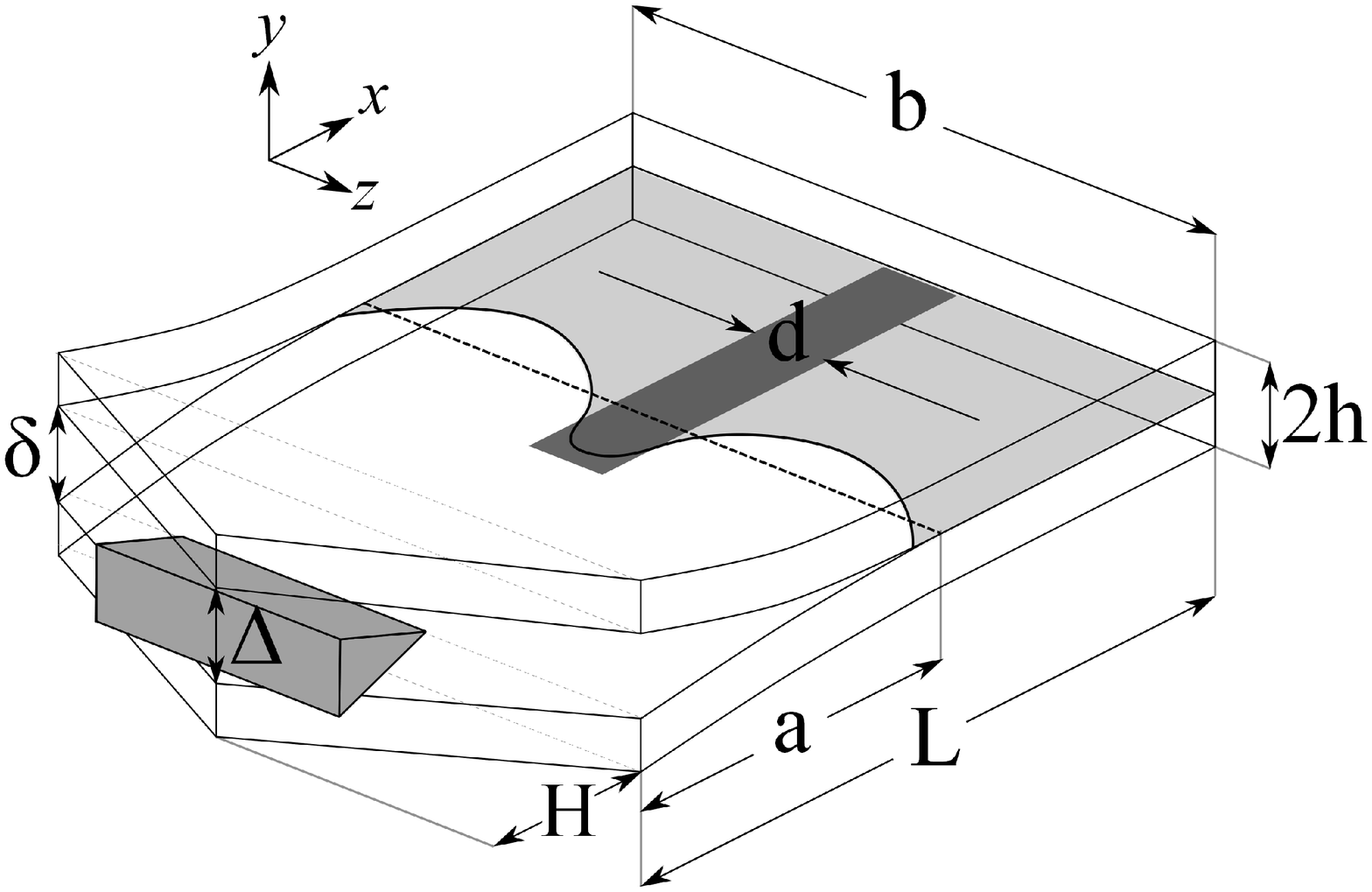}}}$
  $\vcenter{\hbox{\includegraphics[width=0.34\textwidth]{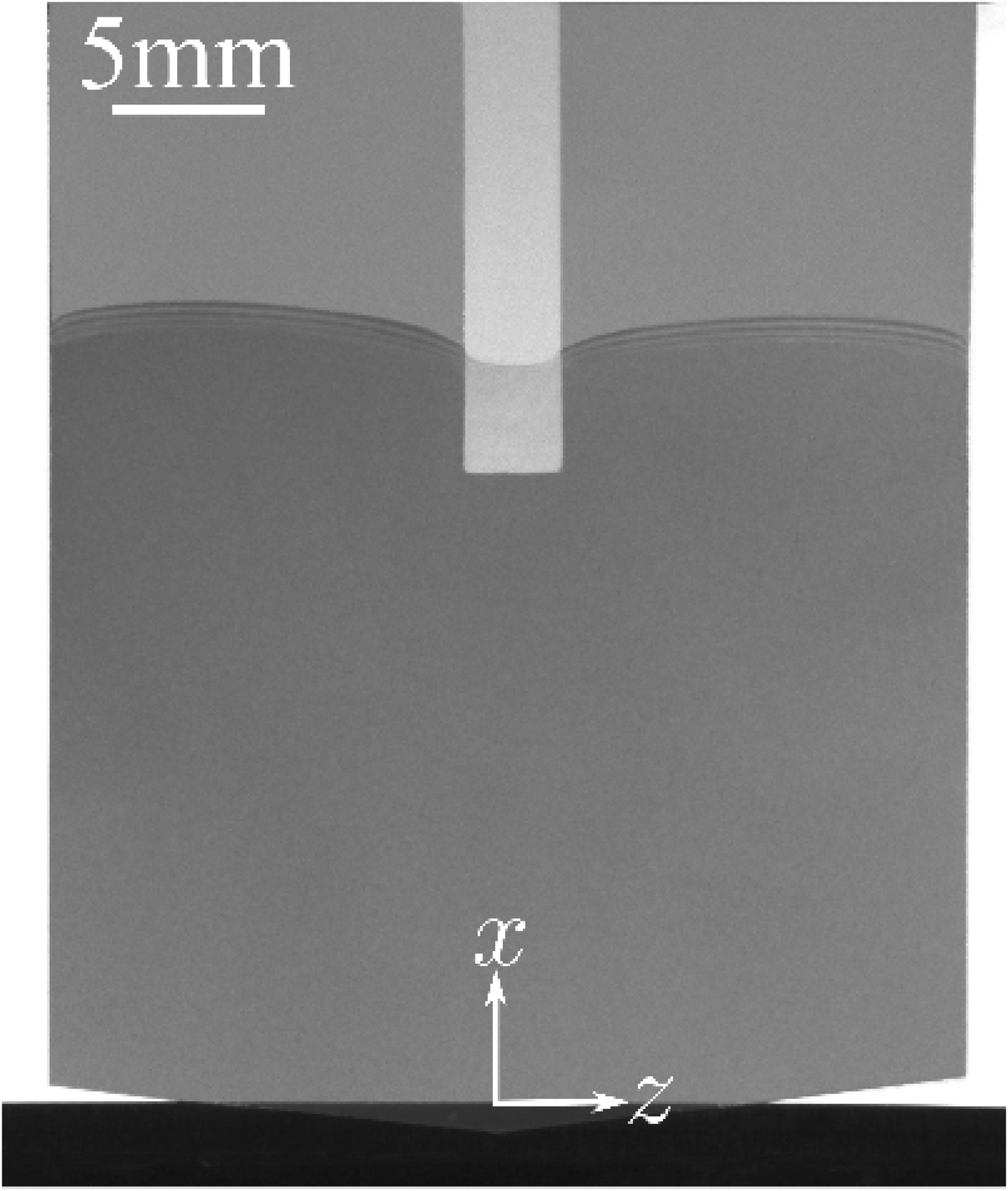}}}$
 \end{minipage}
 \textbf{(a)}
 \hspace{0.55\textwidth}
 \textbf{(b)}
    \caption{(a) Schematic of cleavage test on a double cantilever beam.
The opening of the crack $\delta$ is imposed by the wedge in order to control
the average length of the crack $a$. (b) Top view sample photography during
the cleavage test HC-b30. Thanks to transparency of the glass, interfacial
crack is observed. The deformation of the crack comes from the difference of
adhesion energies between the silver interface and the defect strip.}
\label{fig:load}
\end{figure}

To propagate a crack in the multilayer, we use cleavage under
ambient conditions: a glass backing of thickness 700~$\mu$m is glued on top of
the multilayer with an epoxy glue \citep{Barthel2005,Dalmas2009},
resulting in a glass/multilayer/glass sandwich. This sandwich is
amenable to opening in a DCB setup. The test is carried out by
progressive opening of the two glass arms by the gradual
introduction of a wedge between the two plates
(Fig.~\ref{fig:load}a). The positioning of the wedge is controlled
by an electric jack for precise control of the opening of the two
glass plates. With this method, cleavage is displacement controlled
and the propagation of the crack is stable: the crack length
increases in a controlled way. The propagation of the crack front is
performed in a quasi-static manner, i.e. further increment of the
opening is applied only after complete arrest of the crack front.

The specimen is semi-transparent and we directly observe the crack
front propagation during opening. In order to record the data, two
high magnifications cameras are used. The first camera monitors the
displacement of the glass arms $\delta$, which is measured at the
corners of the crack surfaces (see Fig.~\ref{fig:load}a). The second
camera records the crack front shape $a(z)$ (see Fig.~\ref{fig:load}b). 
Its position $a(z)$ is measured from the line parallel to z passing through 
the external corners of the crack surfaces.

Once the wedge penetrates between the two plates, crack propagation
takes place at the weakest interface of the multilayer. In order to
identify this interface X-Ray photoelectron spectroscopy is used on
both fracture surfaces. In these samples, purely interfacial
fractures have always been found. As indicated in
Fig.~\ref{fig:multilayer}, the crack takes place at one of the
silver layer interfaces for all the configurations. This observation
is consistent with the weak adhesion of silver layers and the
perfectly interfacial crack path demonstrated earlier
\citep{Barthel2005}. As the crack propagates in a patterned
interface, two fracture interfaces can be defined, one involving the
silver layer and the other one in the defect strip (see
Tab.~\ref{ta:stack}).

\begin{figure}[h!]
    \centering
    \includegraphics[width=0.49\textwidth]{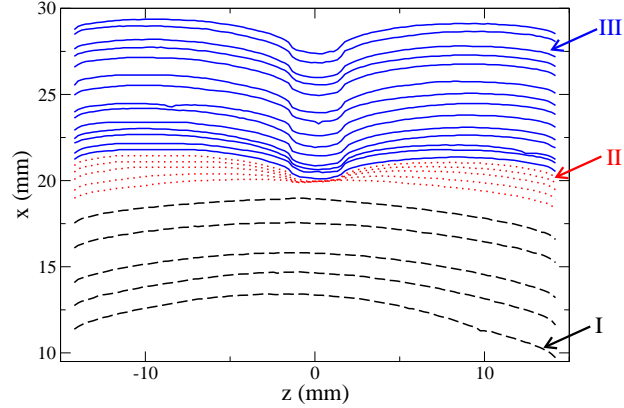}
    \caption{Evolution of the equilibrium shape of the measured crack front for the HC-b30 sample. The three successive regimes of propagation are distinguished. I: Stationary propagation along the silver homogeneous interface (wide dashed black lines); II: transient deformation of the front when entering the defect strip (red dotted lines); III: stationary propagation of the deformed front in the defect strip (continuous blue lines).}
\label{fig:fronts}
\end{figure}

The evolution of the crack front morphologies (only
for equilibrium positions) during cleavage tests is shown in
Fig.~\ref{fig:fronts} in the case of sample HC-b30. In the first
regime, the propagation takes place at the homogeneous silver
interface. The crack front propagates with the overall curved shape
which is characteristic of homogeneous interfaces for finite width
samples (Sec.~\ref{se:FEhomogeneous}). This simple curved shape is
preserved until the crack comes into contact with the defect strip,
where deformation of the crack front begins: it is held back by the
defect strip and curves into the outer region. After this transient
regime, stationary propagation is recovered: in this third regime,
where the crack propagates in the patterned area of the interface,
the overall curvature of the crack front is now decorated by an
additional modulation: the front lags behind in the high adhesion
defect strip.

The crack front positions $a(z)$ and the loadings $\delta$, with the 
sample geometric dimensions and the elastic constants of the glass 
($E$,$\nu$), serve as input parameters to model the cleavage tests 
in the next sections by two different ways. We compute the ERR landscape 
along the crack front first by FE numerical calculations 
(Sec.~\ref{se:FE}), second, by theoretical perturbation approaches 
(Sec.~\ref{se:th}) and compare the results (Sec.~\ref{se:comp}).

\section{ERR obtained numerically by Finite Elements} 
\label{se:FE}

FE calculations are performed with the aim to analyze the validity of the
analytical perturbation approaches proposed in this paper (Sec.~\ref{se:Perturb}). 
To directly model the whole experimental pinning cleavage tests and obtain a reference 
solution of the global and local ERRs, we needed to take into account the whole 
geometry of the problem, including specimen geometry and the crack morphology (Sec.~\ref{se:FEglobal}). 
This is the major part of the present FE calculations (Sec.~\ref{se:G_FE}). Besides, we also consider 
the case of an homogeneous interface (Sec.~\ref{se:FEhomogeneous}). Even in this case, due to the 
finite width of the specimen, the front shapes are not straight and are derived here by FE calculations. 
They are used as reference unperturbed crack geometry for the perturbation methods.

\subsection{Simulation method}
\label{se:FEglobal} 

To calculate the local ERR distribution, the static elasticity problem is resolved by FE for 
each equilibrium position of the crack front. The input
parameters for the FE calculations are the geometrical dimensions of the sample
$(L, b, h, H)$ (Fig.~\ref{fig:load}), along with the position of the front
$a(z)$ and the associated opening $\delta$, which are measured during the
gradual opening of the sample.

\begin{figure}[h!]
    \centering
    \includegraphics[width=0.49\textwidth]{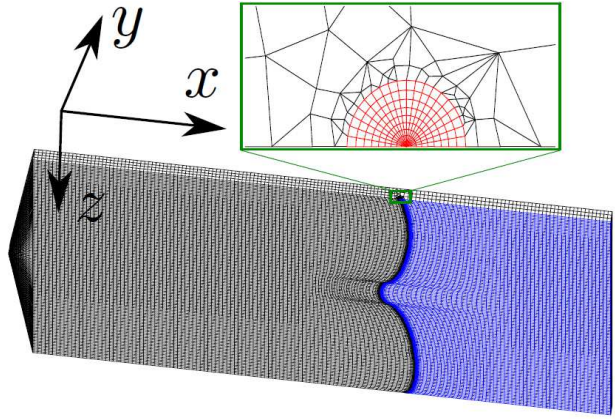}
    \caption{Typical mesh of the cleavage test used in the finite element calculations. 
The unbroken interface is colored in blue. The highest mesh density corresponds to the 
crack front position. A zoom of the mesh in the vicinity of the crack tip is also reported 
in red.}
\label{fig:mesh}
\end{figure}

The simulations are performed with the FE code CAST3M developed by the French
Commissariat \`{a} l'\'{E}nergie Atomique (CEA). The values of Young's modulus
$E=71$ GPa and Poisson's ratio $\nu=0.22$ are those of glass for all the
cleavage tests. The multilayer is neglected from a mechanical point of view due
to its extremely low thickness compared to the glass plates. The results
presented in this work correspond to meshes composed of 144,896 bilinear 8-node
parallelepipedic elements and 180,471 nodes. The mesh of the glass plate is
inhomogeneous. Its density increases as the displacement gradients, and thus
especially in the vicinity of the crack tip. The mesh is refined until the
displacement field converges at mechanical equilibrium. The mesh of the
experimental crack front shape $a(z)=a+\delta a(z)$ is obtained by deforming 
an invariant mesh in the $z$-direction of a straight crack front around its mean
position $a$ (Fig. \ref{fig:mesh}). The mesh around the crack front is
constructed to form a kind of regular cylinder around the front which is
adapted to the ``G-theta'' method used to calculated the local ERR at each
point of the front. This method developed by \cite{Destuynder1981}, allows us
to calculate $G$ accurately and quickly \citep{Destuynder1983}.

As output of the simulation, the 3D displacement fields obtained are used to
obtain the local ERR $G(z)$ along the front with the ``G-theta'' method. This
method uses the displacement fields obtained by FE resolution of the elasticity
problem. Due to the symmetry of the problem with respect to the $x-z$ plane,
i.e. the plane in which the crack propagates, only the upper glass plate
located above the plane $y=0$ is simulated. Symmetry boundary conditions are
applied on this boundary.  The boundary $x=L$ is clamped. Stress free boundary
conditions are applied on the rest of the boundary surfaces. Concerning the
loading, a point displacement $\Delta$ in the $y$-direction is applied on
the triangular tip as in the experiments. The value of $\Delta$ is fixed so
that the displacement at the corner $\delta$ of the sample shown in
Fig.~\ref{fig:load}a converges toward the displacement measured experimentally.
The final error between the experimental and numerical value of $\delta$ is
less than 1\%, i.e. less than the uncertainty of experimental measurement, for
all crack configurations studied.

\subsection{Evolution of the ERR during the propagation along the heterogeneous interface}
\label{se:G_FE}

The local ERRs are calculated along the crack fronts for all equilibrium 
positions of the crack obtained experimentally. In the following we 
illustrate our results in the case of the multilayer HC-b30 that exhibits 
the strongest adhesion contrast among the four types of interfaces presented
in Sec.~\ref{se:exp}.

\begin{figure}[h!]
    \centering
    \textbf{(a)}
    \includegraphics[width=0.46\textwidth]{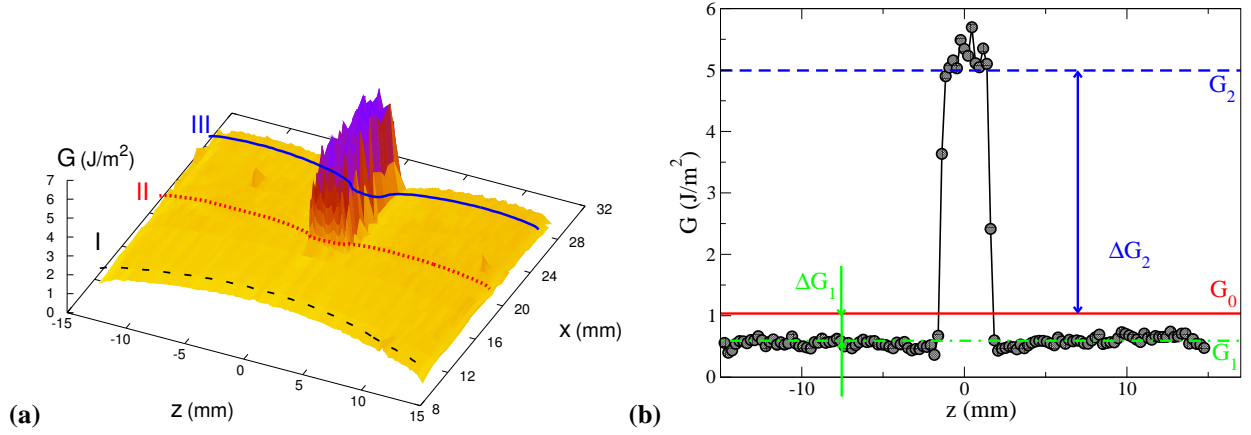}
    \textbf{(b)}
    \includegraphics[width=0.46\textwidth]{5b.eps}
    \caption{Local Energy Release Rate (ERR) computed by Finite 
Element (FE) for the cleavage test HC-b30.
(a) Evolution of the ERR landscape. The three successive stages of propagation 
are distinguished as in Fig. \ref{fig:fronts}. The ERR contrast expected 
between the defect strip and its surrounding homogeneous medium is well 
reproduced by the ERR landscape.
(b) Local ERR along a crack front during regime III of Fig.~\ref{fig:fronts}. 
$G_0$ denotes the average along the front while $\Delta G_1$ and $\Delta G_2$ 
denote the average variations of the energy release rate compared to $G_0$ 
outside and inside the defect strip respectively.}
\label{fig:G_z}
\end{figure}

\begin{figure}[h!]
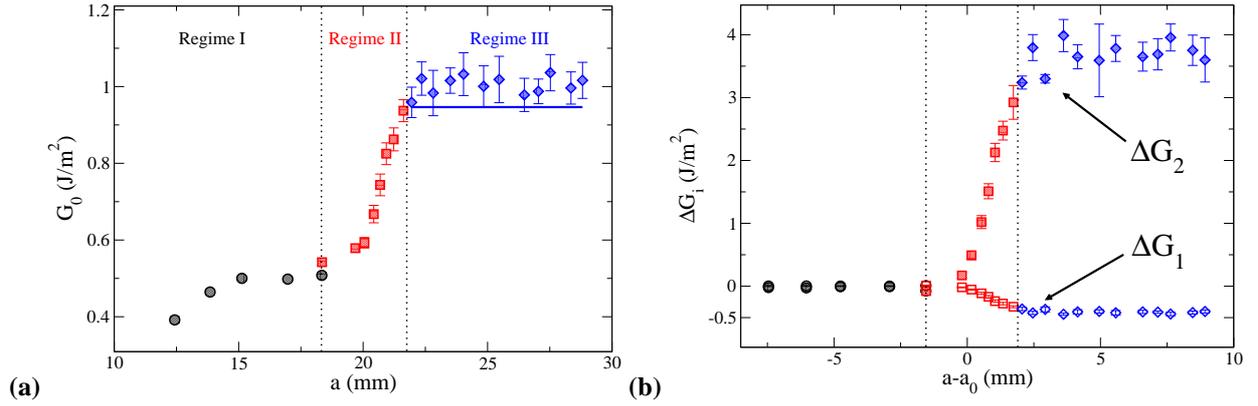

    \centering
    \textbf{(a)}
    \includegraphics[width=0.46\textwidth]{6a.eps}
    \textbf{(b)}
    \includegraphics[width=0.46\textwidth]{6b.eps}
    \caption{Energy Release Rate (ERR) calculated by finite element 
method (symbols) for the cleavage test HC-b30. In order to distinguish 
the three successive regimes of propagation the color code is same as 
in Fig.~\ref{fig:fronts}: black circles for the propagation along 
the homogeneous part of the interface, red squares for the transient 
regime and blue diamonds for the propagation in the patterned part of 
the interface. The error bars are deduced from statistical dispersions. (a) Mean ERR $G_{0}$ evolution with
the mean crack front position $a$. The continuous straight line corresponds
to the theoretical value deduced from the linear regression of 
Fig.~\ref{fig:Kanninen} computed through Eq.~\ref{eq:Kanninen}.
(b) ERR contrasts $\Delta G_{i}$ as a function of the distance between the average
crack front position and the beginning of the defect strip position $a-a_{0}$. The
contrasts are calculated outside ($\Delta G_{1}$) and inside ($\Delta G_{2}$) the
defect strip.}
\label{fig:delta_G}
\end{figure}

We plot the following results:
\begin{itemize}
\item The ERR landscape $G(z)$ of a crack front trapped by the central
  defect strip (Fig. \ref{fig:G_z});
\item The average ERR $G_0=<G(z)>$ as a function of the average
  position $a$ of the crack front (Fig. \ref{fig:delta_G}a);
\item The average fluctuations $\Delta G_1$ and $\Delta G_2$ of $G(z)$
  with respect to $G_0$ for interface areas situated outside and
  inside the central defect strip respectively (Fig. \ref{fig:delta_G}b).
\end{itemize}

The evolution of the ERR landscape $G(z)$ when the crack propagates 
is reported in Fig. \ref{fig:G_z}a. The ERR contrast expected between the defect 
strip and its surrounding homogeneous medium is well reproduced by the ERR landscape. 
As reported in Fig.~\ref{fig:fronts}, three successive stages can be associated with 
this ERR landscape: homogeneous propagation (the front has a curved shape characteristic 
of finite width effect); interaction between the crack front and
the central defect strip (the front is progressively deformed); stationary pinning (the deformation 
stays invariant in the direction of propagation).

A profile of the ERR $G(z)$ obtained for a crack front trapped by the
defect strip in the last stationary regime is shown in Fig. \ref{fig:G_z}b. 
Despite the presence of some fluctuations, the presence of the central defect 
strip can be clearly identified and characterized by a significant variation 
of the ERR compared to average ERR $G_0$. In the homogeneous
and the defect zones, the ERR is nearly constant longitudinally to the
crack front in the direction $z$. It is therefore possible to
attribute a characteristic ERR value for the two types of interfaces,
homogeneous and defect strip, encountered by the crack. We note $G_1$ and
$G_2$ their respective average value from which we extract the
corresponding mean variations from $\Delta G_1=G_1-G_0$ and
$\Delta G_2=G_2-G_0$. The signs of $\Delta G_1$ and $\Delta G_2$ are
by definition opposed. The notations $G_0$, $G_1$, $G_2$, $\Delta G_1$
and $\Delta G_2$ are reported on Fig. \ref{fig:G_z}b.

Thanks to the FE calculations we compute the global ERR $G_0$ by calculating  
the average ERR along the front knowing the local ERRs so that $G_0=<G(z)>$.
In Fig. \ref{fig:delta_G}a, the variation of $G_0$ shows the three regimes as a 
function of the average position of the crack in its direction of 
propagation $a$. These regimes correspond again to the 
morphology variations reported in Fig. \ref{fig:fronts}. The average 
ERR $G_0$ is almost constant at the beginning and at the end of cleavage
test. Between these two ERR values a transient regime is
observed. These three regimes correspond chronologically to the
propagation of the crack front in a homogeneous medium, its
interaction with the strip edges and finally to its
stationary pinning by the defect strip. Changes in $G_0$ reflect the change
in the effective toughness of the interface due to the presence of the
defect strip during the crack propagation. The increase in $G_0$ can
be associated with a toughening of the interface. This aspect will be discussed 
in more detail in Sec.~\ref{se:properties}.

The determination of local ERR values allows us to refine our understanding of the
interaction between the crack front and the central defect strip. In
Fig. \ref{fig:delta_G}b, we represent the variations $\Delta G_1$ and
$\Delta G_2$ as a function of the distance between the average
position of the crack $a$ and the beginning of the masked zone $a_0$ in
the crack propagation direction $x$. The three regimes described
above, from the change in $G_0$, are considerably emphasized. We
observe that the variations $\Delta G_i$ are equal to zero for
$a-a_0<0$, i.e. before the interaction of the crack with the
defect. During the penetration of the strip by the crack front, in the
vicinity of $a-a_0=0$, $\Delta G_i$ vary dramatically. The $\Delta
G_i$ finally reach constant values for $a-a_0>0$ when the crack front
is stationary trapped by the defect. Note that, in absolute value, the
variation $\Delta G_1$ is about one order of magnitude lower than
$\Delta G_2$. The width of the homogeneous zone is indeed an order of
magnitude greater than that the defect strip one.
The signs of variations $\Delta G_i$ are consistent with the observed variations 
in $G_0$. A defect strip whose adhesion is larger (smaller) than 
the homogeneous medium produces a variation $\Delta G_2>0$ ($\Delta G_2<0$) 
and $\Delta G_1<0$ ($\Delta G_1>0$). 

\subsection{Simulation of the overall crack front curvature for an homogeneous interface}
\label{se:FEhomogeneous}

Even for a homogeneous pattern-free interface, for a
sample of finite width $b$, the crack fronts are not exactly straight but adopt
a curved shape (see wide dashed black lines in Fig.~\ref{fig:fronts}). This shape 
is due to the anticlastic deformation of the bent plates. To our knowledge, there 
is no simple analytical formula for modeling this contribution to the crack front 
shape. Therefore we have calculated the crack shape in an equivalent homogeneous 
medium corresponding to the experimental geometry and loading. Using the FE method 
we have calculated crack shapes such that $G(z)=G_0$ along the front where $G_0$ 
is the average ERR computed previously by FE. The crack front shape is determined 
from a mere steepest descent iterative algorithm. The equilibrium shape is reached 
when the difference between the local ERR $G(z)$ and $G_0$ is less than 1\% at any point 
of the front. These background curved shapes will then be subtracted from the experimental 
fronts before analysis of the deformed crack fronts in the heterogeneous samples with 
the perturbation formulas (Sec.~\ref{se:comp}).

\section{Analytical approaches}
\label{se:th}

The idea here is to decompose the calculation of the local ERR along the deformed crack front 
as the sum of its global average value and its local increase along the perturbed configuration. 
Analytical methods allowing us to estimate the mean value (beam model) and the local relative values 
(perturbation methods) are presented in sections \ref{sec:beam} and \ref{se:Perturb} respectively. Our 
goal is here to be able to determine the ERR landscape analytically in order to compare theoretical 
predictions with previous FE numerical calculations in sections \ref{se:comp} and \ref{se:properties}.

\subsection{Determination of the average ERR}
\label{sec:beam} 

The first step of our theoretical study is to determine the global 
ERR, i.e. the mean ERR $G_{0}$. A standard data analysis is to calculate 
$G_{0}$ by a simple beam bending model at fixed displacement (Fig.~\ref{fig:load}b). 
According to~\cite{Kanninen1973}, the ERR in a state of plane stress reads:
\begin{equation}
\label{eq:Kanninen}
G_{0}=\frac{3Eh^{3}\delta^{2}}{16(a+0.64h)^4}.
\end{equation}
Note that the geometry of the problem meets the assumptions leading
to Eq.~\ref{eq:Kanninen}, i.e. $a \gg h$ and $L-a>2h$, for all the
cleavage tests studied in this work. We need to adapt the beam model to pass from 
a three-dimensional elastic problem to a beam model in one dimension. Kanninen's model is 
therefore employed using the average crack front position $a$ calculated 
along the width of the plate in the $z$ direction. This average position 
is obtained directly from experimental crack front positions $a(z)$. 

\begin{figure}[h!]
    \centering
    \includegraphics[width=0.5\textwidth]{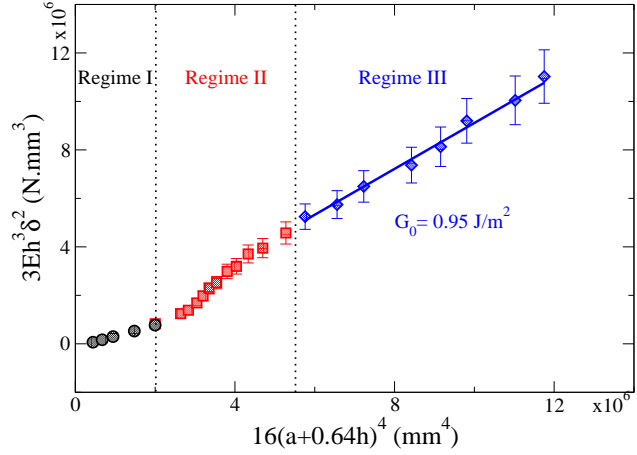}
    \caption{Typical data plot according to Kanninen's model for 
the cleavage test HC-b30. The slope corresponds to the mean Energy Release Rate (ERR) $G_0$.
The three successive regimes of propagation are distinguished as 
in Fig.~\ref{fig:fronts}. I: stationary propagation along the silver
homogeneous interface (black circles); II: transient deformation of the 
front when entering the defect strip (red squares); III: stationary propagation 
of the deformed front in the defect strip (blue diamonds). The straight line 
is a linear regression of the data in the regime III.}
\label{fig:Kanninen}
\end{figure}

In practice, we plot the evolution of $\delta^{2}$ as a function of
$(a+0.64h)^4$. In Fig.~\ref{fig:Kanninen}, the case of sample HC-b30 is
reported as an example. In this kind of graph, $G_{0}$ is simply proportional 
to the slope of the curve. We can clearly identified the three different regimes 
described in Sec.~\ref{se:G_FE}. 

We extract the nearly constant value of $G_{0}$ in regime III, i.e. where the the pinning is 
stationary, by a linear regression procedure. In the case of 
sample HC-b30, we obtain $G_{0}= 0.95\pm0.14$~J/m$^2$  as reported in 
Fig.~\ref{fig:Kanninen}. This quantity will be used to determine adhesion 
properties in Sec.~\ref{se:properties}. The comparison with FE 
numerical calculations in Fig.~\ref{fig:delta_G}a shows a good agreement. 

For all the experiments, the variations of $G_0$ due to the interaction between the crack 
and the defect are well reproduced by Eq.~\ref{eq:Kanninen} as a function of the average 
position of the crack front. $G_0$ is only slightly underestimated with respect to FE 
calculations. The maximum relative error between the beam model predictions and the FE 
numerical calculations remains below $10$\% for all samples. This result 
validates the use of Kanninen's model to evaluate $G_0$ in our system.

\subsection{Perturbation method}
\label{se:Perturb}

We now present the perturbation approaches to model the cleavage test. Our goal 
is to evaluate the local ERRs from the knowledge of the crack front morphology 
$a(z)$, this time using a lightweight analytical method. We first discuss the 
formulas giving the first order variation of the local ERR due to a perturbation 
of the crack front position. We then customize them to our experiments: we assume that 
the ERR landscape is stepwise to reduce the determination of the local ERR to a single parameter, 
namely the ERR contrast. The usefulness of those formulas appears when applied to the
experiments, since they allow to reduce the errors linked to the local noise of measures 
in the crack front position $a(z)$. 

\subsubsection{Variation of the local ERR due to small crack front perturbations}

Two different cases are considered: 1) when the wavelength $\lambda$ of the crack 
front perturbation is small compared to the plate thickness, that is when the medium 
can be supposed to be infinite and 2) when $\lambda$ is large, that is when a thin 
plate model can be used.

$\bullet$~\textbf{\cite{Rice1985}'s formula for a half-plane crack in an infinite medium}:
In his pioneer work \cite{Rice1985} has calculated the theoretical expression 
of the ERR first-order variation $\delta G(z)$ along a planar crack front due to a small perturbation 
of the front position $\delta a(z)$. This expression is valid in principle for a
semi-infinite crack in an infinite body. From the perspective of a DCB geometry
(see Fig.~\ref{fig:load}a), such a geometric mapping is acceptable if the
characteristic perturbation length $\lambda$ of the crack front meets the
conditions: $\lambda \ll h$ and $\lambda \ll a$. The first condition implies
that the plate must be very thick with respect to the perturbation wavelength.
Provided that these assumptions are met and that the load is invariant along
the crack front in the $z$ direction, $\widehat{\delta G}(k)$ is given in
Fourier space by:
\begin{equation}
\frac{\widehat{\delta G}(k)}{G_0}=F_{\infty}(k,a)\widehat{\delta a}(k) \qquad \textrm{with} \qquad F_{\infty}(k,a)=(-|k|+\frac{1}{G_0}\frac{dG}{da}),
\label{eq:delta_GRice}
\end{equation}
where $\lambda=2\pi/|k|$ is the wavelength of the front perturbation
and $G_0$ the ERR of the straight crack front of average position
$a$. Note that to first order $G_0$ corresponds to the average ERR as 
introduced in the previous sections. The derivative $dG/da$ is calculated 
in our work from the simple Euler-Bernoulli beam model \citep{Tada1985} 
consistent with the plate model presented subsequently. In this case, the 
variation $1/G_0.(dG/da)$ with the crack length $a$ is merely given by $-4/a$.\\\\

$\bullet$~\textbf{\cite{Legrand2011}'s formula for a half-plane crack in a thin plate}\\
Rice's method has been extended by \cite{Legrand2011} adapting it to the case
of a semi-infinite planar crack located in the mid-plane of a thin plate. This
system satisfies the conditions: $\lambda \gg h$ and $h \ll a$. Under these
assumptions, the perturbation of the crack can be treated with the
Love-Kirchhoff theory for thin plates, which greatly reduces the complexity of
the problem. Note that in this case, the first condition implies that the plate
must be very thin with respect to the perturbation wavelength. The first order variation 
of the ERR in Fourier space writes as follows:
\begin{equation}
\frac{\widehat{\delta G}(k)}{G_0}=F_{Plate}(ka)\widehat{\delta a}(k) \qquad \textrm{with} \qquad F_{plate}(ka)=\frac{2}{a}\left(\frac{2ka\cosh(2ka)-\sinh(2ka)}{2ka-\sinh(2ka)}\right).
\label{eq:delta_Gplaque}
\end{equation}

\subsubsection{Application to a stepwise ERR landscape}

The cleavage experiments can be modeled by assigning characteristic 
ERRs to the homogeneous area ($G_1$) and defect strip ($G_2$). This 
assumption seems very reasonable from FE numerical results of 
Sec.~\ref{se:G_FE}. In the presence of a defect \emph{of known width} $d$, 
the ERR landscape is modeled by:
\begin{equation}
\label{eq:Gland}
G(z)=G_0+\Pi_1(z)\Delta G_1+\Pi_2(z)\Delta G_2 \qquad \textrm{with} \qquad
\left\lbrace
\begin{array}{cc}
\Pi_1=0, \Pi_2=1 & \textrm{if } |z|<d/2 \\
\Pi_1=1, \Pi_2=0 & \textrm{if } |z|>d/2
\end{array}\right. ,
\end{equation}
where $\Pi_i$ are rectangular functions. The variation of the normalized ERR is:
\begin{equation}
\label{eq:Gvarland}
\frac{\delta G(z)}{G_0}=\frac{\Delta G_1}{G_0}\Pi_1(z)+\frac{\Delta G_2}{G_0}\Pi_2(z).
\end{equation}
The average $G_0$ provides a relation linking the ERR of the defect
strip and the surrounding homogeneous medium. Denoting $G_1$ the ERR 
in the homogeneous zone and $G_2$ the ERR in the defect strip, we assume 
that $G_0$ is an average along the front line, which we simply write as:
\begin{equation}
\label{eq:Gmoy}
G_0=\frac{G_1 (b-d)+G_2 d}{b}.
\end{equation}

We can write Eq.~\ref{eq:Gvarland} in Fourier space to deduce the
front shape $\delta a(z)$ from Eqs.~\ref{eq:delta_GRice} or 
\ref{eq:delta_Gplaque}. Using Eq.~\ref{eq:Gmoy} to express
$\Delta  G_2$ as a function of $\Delta G_1$, we obtain:
\begin{equation}
\label{eq:final} \widehat{\delta a}(k)=\frac{1}{F(k,a)}\frac{\Delta
G_2}{G_0}\left[-d/(b-d)\widehat{\Pi_1}(k)+\widehat{\Pi_2}(k)\right],
\end{equation}
where $F(k,a)$ is one of the elastic kernels from
Eqs.~\ref{eq:delta_GRice} or \ref{eq:delta_Gplaque}.
Eq.~\ref{eq:final} establishes a relationship which relates the
shape of the crack front with the ERR contrasts between the defect
strip and the homogeneous medium in the present geometry.

\section{Determination of ERR contrasts}
\label{se:comp}

In the previous sections, we have presented experimental front
morphologies in homogeneous and patterned interfaces, along with
thorough FE modeling to access both global ERR and ERR distribution
over the sample. We now analyze the same data with the analytical
formulas derived from the perturbation method.

\subsection{Corrections of experimental crack front positions}

In order to apply the perturbation approaches to calculate the local
ERR, it is first necessary to correct the experimental front shapes
$a(z)_{Exp}=a+\delta a(z)_{Exp}$ from extrinsic effects that were not 
taken into account in the perturbation method. This task is equivalent 
to determining the \emph{unperturbed} crack front geometry and its 
variation with crack length $\delta a(z)$ in the absence of defect strip. 
Indeed, the perturbation approaches described in Sec.~\ref{se:Perturb}
apply only to a solid which is infinite in the lateral ($z$)
direction, \emph{i.e.} an infinite straight crack front. In the
experiments, the finite width of the sample induces a curvature of
the front due to the anticlastic effect as presented in
Sec.~\ref{se:FEhomogeneous}. In order to take into account this
additional finite size effect, we take as unperturbed reference
crack front, for a given global $G_0$ level, the curved front
$a(z)_{Anti}=a+\delta a(z)_{Anti}$ calculated by the FE simulations 
method presented in Sec.~\ref{se:FEhomogeneous}.

\begin{figure}[h!]
    \centering
    \includegraphics[width=0.5\textwidth]{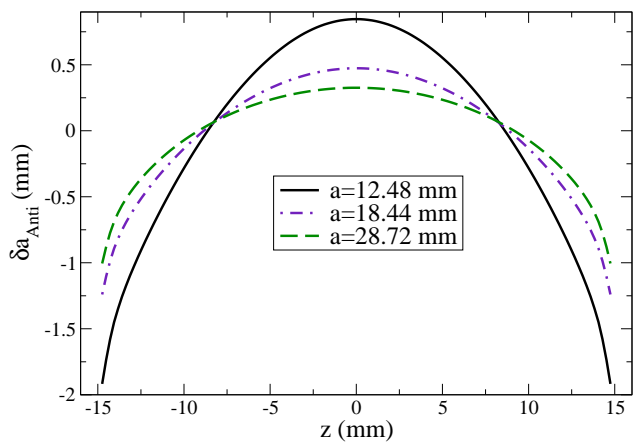}
    \caption{Variation of the crack front position calculated by the finite element
method for an equivalent homogeneous medium whose curvature comes from anticlastic
coupling. The obtained shapes are exemplified for the HC-b30 cleavage test for
different mean crack front positions. Note that the scale of the axes are different. 
The front is almost straight.}
\label{fig:homogeneous}
\end{figure}

Examples of curved fronts obtained by this method for the HC-b30 cleavage 
test are shown in Fig.~\ref{fig:homogeneous}. These shapes are characteristic 
of curved crack fronts in homogeneous specimens of finite width \citep{Jumel2008}. 
Note that, due to the changes of the ratio $a/b$, and thus the stress state with 
the advance of the crack, the curvature of the crack depends on the average crack 
length as expected. The loading on the triangular tip of the plates also reinforces the crack 
front curvature variation. This curvature variation shown in Fig.~\ref{fig:homogeneous} emphasizes 
that the influence of the width of samples is not constant as assumed in \cite{Dalmas2009}. 
It clearly depends on the the mean crack length $a$. The solutions obtained will serve us as 
the reference unperturbed solutions in the perturbation approaches presented below.

A second correction comes from a slight rotation around the x-axis
of the wedge between the plates during loading which induces an
imperfect symmetry with respect to the plane $z=0$ at the center of
the sample. This lack of symmetry is reflected by a slight linear
bias in $z$ of the crack position. For instance, the crack front
shape, corrected from the  anticlastic effect, in a homogeneous
medium is a linear function of z instead of being completely flat
with $\delta a(z)=0$ for all points along the front. We choose to
correct this load-related effect by subtracting a linear fit $\delta
a(z)_{Lin}$ to the experimental position. This procedure partially
restores the symmetry of the problem with respect to the plane z=0
at the center of the specimen as assumed in the model described in
the previous theoretical section. Note that it does not affect the
ERR contrast calculation, but merely allows a better fit of the
elastic line model on experimental data.

The variation of the position of the crack front around its average 
position $a$ corrected from both anticlastic and rotation effects writes:
\begin{equation}
\label{eq:correction_position}
\delta a(z)=\delta a(z)_{Exp}-\delta a(z)_{Anti}-\delta a(z)_{Lin}.
\end{equation}

\begin{figure}[h!]
    \centering
    \includegraphics[width=0.5\textwidth]{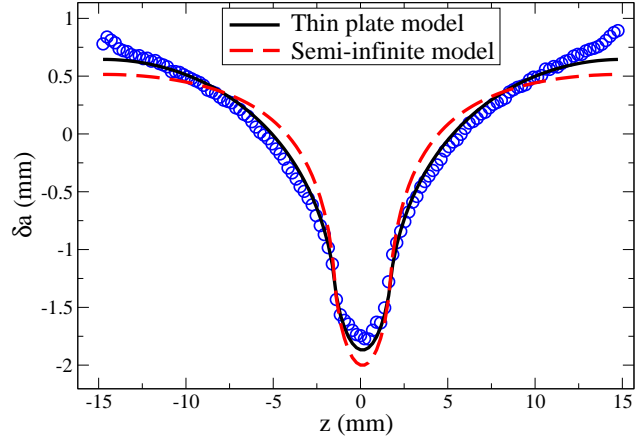}
    \caption{Variation of the experimental crack front positions (symbols) 
corrected through Eq.~\ref{eq:correction_position} for the cleavage test HC-b30. 
The dashed and solid curves correspond to the fitted semi-infinite 
(Eq.~\ref{eq:delta_GRice}) and thin plate (Eq.~\ref{eq:delta_Gplaque}) models, 
respectively. Note that the vertical axis is expanded approximatively ten times 
compared to the horizontal axis.}
\label{fig:fitexp}
\end{figure}

An example of a corrected crack front morphology in regime III for
the HC-b30 sample is shown in Fig.~\ref{fig:fitexp}. The impact of
the defect strip on the front morphology is clearly visible.
Consistent with the FE analysis of Fig.~\ref{fig:G_z} and
~\ref{fig:delta_G}, an increase in ERR corresponds to the anchoring
of the crack. Note that the ordinate axis of Fig.~\ref{fig:fitexp}
is approximatively expanded ten times in comparison with the axis of
abscissa. This observation reflects the fact that the perturbations
of the crack front position are small compared to the width of the
specimen. The morphologies of experimental crack fronts therefore
justify a priori a theoretical treatment based on a first-order
perturbative approach presented in Sec. \ref{se:Perturb}.

\subsection{Quantitative analysis of crack front deformations with perturbation methods}
\label{se:quant}

In order to reproduce the shape of the experimental crack front with
Eq.~\ref{eq:final}, we have chosen to work in the direct space by
adjusting the single free parameter $\Delta G_{2}/G_{0}$. Indeed,
the presence of experimental noise on the crack front position makes
the determination of the ERR contrast difficult from a direct use of
Eq.~\ref{eq:final} in Fourier space.  The normalized ERR
contrast is determined by minimizing the mean square error between
the experimental front shape and the front morphology predicted
after a numerical transformation into real space of
Eq.~\ref{eq:final}. This adjustment procedure has fewer degrees of
freedom than in the work of \cite{Dalmas2009} where the width of the
defect strip was taken as a free parameter.

The morphologies of the fronts obtained from the best fit of the
different elastic kernels are plotted in Fig.~\ref{fig:fitexp}. Both
perturbative approaches satisfactorily reproduce the crack front
shapes except near the edges of the specimen due to free surface
effects. Fig.~\ref{fig:fitexp} shows only a slight better agreement
with the experimental results for the thin plate model. In order to
get quantitative interpretation, the average mean square error over
all the crack configurations for the four samples has been computed
for both models. We found that the error on the experimental crack
front positions is lower in average for the thin plate model than
for the semi-infinite model. It seems difficult however to find a
significant difference between the semi-infinite and thin plate
formulas on the basis of the front morphology. In the next section
we will see that quasi-equivalent front shapes in fact involve very
different ERR contrasts allowing us to judge of the applicability
of the models.

\subsection{Comparison between perturbation methods and FE results}

The fitting procedure presented above gives us the opportunity to
access local ERR contrasts, i.e. to the difference of ERR between
the defect strip and its homogeneous environment. From this
adjustment we can calculate the normalized contrasts $\Delta
G_2/G_0$ directly through Eq.~\ref{eq:final}. $\Delta G_1/G_0$ is then
deduced from $\Delta G_2/G_0$ through Eq.~\ref{eq:Gmoy}. This last step
completes the determination of the ERR landscape by the perturbation
method, knowing the geometry of the system, especially the crack
front morphology and the defect strip width.

As expected from the theoretical work of \cite{Legrand2011}, the
absolute values of ERR contrasts predicted by the semi-infinite
model are found to be lower than those predicted by the thin plate
model. The ratio between the latter and the former is about 2.5. To
discriminate between the two elastic line models, we compare their
respective predictions with the numerical FE calculations.

\begin{figure}[h!]
    \centering
    \includegraphics[width=0.5\textwidth]{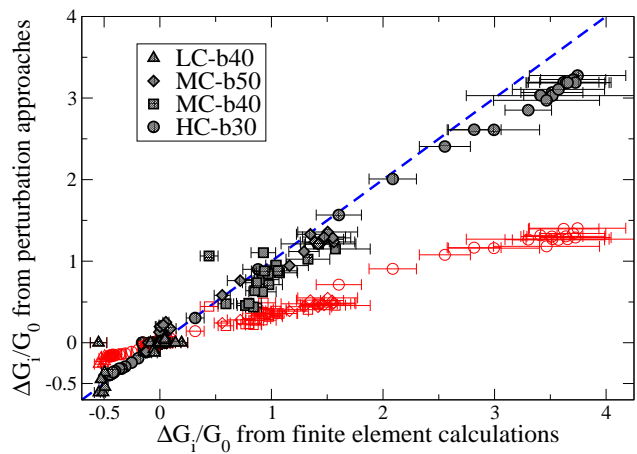}
    \caption{Comparison of normalized Energy Release Rate (ERR)
contrasts $\Delta G_i/G_0$ calculated by the finite element method 
with those obtained from the semi-infinite (red open symbols) and the thin plate 
(black solid symbols) theoretical approaches. The blue dotted lines correspond to the 
perfect equivalence between these two quantities. The error bars are deduced from 
statistical dispersions.}
\label{fig:compa}
\end{figure}

Comparisons between the ERR contrasts deduced from theoretical
approaches with the FE results are shown in Fig.~\ref{fig:compa}
where the results obtained for all crack fronts of the four samples
LC-b40, MC-b40, MC-b50, and HC-b30. Despite some discrepancies,
Fig.~\ref{fig:compa} shows a much better description of the ERR
contrasts when the plate model is used. The average relative error
on $\Delta G_i/G_0$ for all the data of Fig.~\ref{fig:compa} in
comparison with the FE results decreases from 61\% when Rice's model
is applied directly as for a semi-infinite medium to 16\% when the
finite thickness of the plate is taken into account.

Despite a correct description of the  front shapes of the pinned
cracks, Rice's model as applied directly to the present DCB tests systematically 
underestimates the ERR normalized contrasts. 
This result comes from the fact that the perturbation wavelengths of 
the crack front are much larger than the specimen thicknesses. This 
is often encountered in the literature when thin plate geometries is
used to study interfacial crack propagation. Therefore, in such
geometries, it is necessary to model the cracks  by taking into
account the finite thickness of the plate, as shown by
\cite{Legrand2011}. Using this model, we find quantitative
agreement, \emph{i.e.} within the error bars, with the three
dimensional FE calculations.

In the more general case, \cite{Legrand2011} have also developed a
model that connects both regimes (semi-infinite and thin plate) as a
function of the wavelength of the perturbations relative to the
thickness of the specimen. This more general kernel is not relevant
here since the width of the defect is much larger than the plate
thickness $h$, so that we always stand in the vicinity of the thin 
plate limit.

\section{Crack front propagation and material properties}
\label{se:properties}

\subsection{Propagation criterion}

Thanks to our experimental procedure, the interpretation of 
the previous results in terms of material properties, i.e. adhesion, 
is fairly straightforward. Indeed, due to the quasi-static propagation 
of the crack, a propagation criterion has to be fulfilled at any point 
of the observed front. For a given load, the crack front stops at locations 
where the propagation driving force is no longer sufficient. 
It is common for brittle fracture to assume that the crack front advance 
is ruled by the Griffith criterion. In this framework, the equilibrium morphology 
of the crack front $a(z)=a+\delta a(z)$ for a given opening $\delta$ obeys:
\begin{equation}
G(z) \leq G_c(z),
\label{eq:Griffith}
\end{equation}
where $G_c(z)$ is the local adhesion energy. Therefore, we can 
deduce the fracture energies of both the surrounding homogeneous medium and 
the defect strip interfaces denoted $G_{1c}$ and $G_{2c}$ respectively. In 
a very similar way, the adhesion energy has been measured for 
homogeneous interfaces by \cite{Barthel2005}.

While the data of the crack front geometry and the loading geometry 
can give a direct access to the ERR $G(z)$, the same is not systematically 
true for the adhesion energy $G_c(z)$. It would be the case if the 
arrest condition for a crack front Eq.~\ref{eq:Griffith} was an equality 
and not only a simple inequality. Thus, the fracture energy can only be deduced 
from the ERR for stationary crack propagation regimes, i.e. where the materials 
properties are invariant in the crack propagation direction \citep{Roux2003}. In our system, 
it is the case for the regimes I and III described previously. Between these two 
stationary regimes a transient regime (II) is observed where the inequality of 
Eq.~\ref{eq:Griffith} remains and the determination of
$G_c(z)$ is not possible.

In the first regime (stage I of Fig.~\ref{fig:fronts}), the crack
propagates in a completely homogeneous zone and $G_1=G_{1c}$ along
the front. When the crack begins to interact with the obstacle
(stage II of Fig.~\ref{fig:fronts}), $G_2\leq G_{2c}$ in the defect
strip and $G_2$ increases with the applied load involving a
significant change in $\Delta G_i$. Once the defect is penetrated by
the crack $G_2= G_{2c}$ in strip and, for each increment of the
opening, the crack advances with $\Delta G_i$ nearly constant. In
this last stationary regime of crack propagation (stage III in Fig.
\ref{fig:fronts}) the toughness contrast is invariant in the
direction of propagation, and we have  equality
between the local values of adhesion energy and ERR all along the front. The present methods 
can then be used to measure interfacial toughness in the defect strip.

Note that it is also possible to simulate the quasi-static 
propagation of a crack for a given toughness landscape. The adhesion energy of 
the defect strip and its surrounding homogeneous medium could be obtained 
by adjusting their values in order to reproduce the shape of the crack front observed 
experimentally. This procedure is however computationally demanding as 
it implies to describe the propagation of the crack which involves the calculation 
of a large number of configurations. In this study, the aim was not to perform propagation 
simulations of the crack, but to validate the analytical perturbation approach, hence it was enough 
to take advantage of our experimental setup which offers a direct observation of the crack front 
equilibrium shape and to perform direct simulations of the elastic problem for each equilibrium 
configuration.

\subsection{Determination of the defect strip adhesion}
\label{se:adh} 

In this last section we leverage the originality of this work by focusing 
on the determination of the defect strip adhesion $G_{2c}$ in the 
framework  of crack pinning in heterogeneous interfaces. The measurement 
of $G_{2c}$ is carried out in the regime III where the crack front pinning
is invariant in the direction of propagation.

In this regime, we compute the fracture energy $G_{2c}$ as the sum
of global ERR plus the average ERR fluctuation in the defect strip
$G_0+\Delta G_2$. Since $G_{2c}$ fluctuates due
to experimental noise we choose to compute its average
$\overline{G_{2c}}$ over all the crack front configurations
belonging to the last stationary pinning regime (blue diamond
symbols in Fig.~\ref{fig:delta_G}). In the case of FE numerical 
calculations, $\overline{G_{2c}}$ is simply computed from the mean local 
ERRs $G(z)$ in the defect strip as presented in Sec. \ref{se:G_FE}. In the case of
 analytical approaches, $G_0$ is computed with Eq.~\ref{eq:Kanninen} 
(Sec. \ref{sec:beam}) while $\Delta G_2$ are determined through 
the fit of the experimental crack front positions by the two perturbation 
approaches (Sec. \ref{se:quant}). We finally computed $\overline{G_{2c}}$
by averaging $G_0+\Delta G_2$.
 
\begin{table}
    \centering
        \begin{tabular}{|c|c|c|c|c|}
        \hline
Sample & Stack & Finite element (J.m$^{-2}$) & Rice model (J.m$^{-2}$) & Plate model (J.m$^{-2}$) \\ \hline
LC-b40 & Gl./Org/Ag/Si$_3$N$_4$ & 0.35 $\pm$ 0.04   & 0.46 $\pm$ 0.19 &  0.32 $\pm$ 0.13\\ \hline
MC-b40 & Gl./Si$_3$N$_4$/Ag/Si$_3$N$_4$ & 1.57 $\pm$ 0.11   & 1.13 $\pm$ 0.09 &  1.52 $\pm$ 0.14\\ \hline
MC-b50 & Gl./Si$_3$N$_4$/Ag/Si$_3$N$_4$ & 1.86 $\pm$ 0.12   & 1.16 $\pm$ 0.12 &  1.75 $\pm$ 0.18\\ \hline
HC-b30 & Gl./Ag/Si$_3$N$_4$ & 4.63 $\pm$ 0.31 & 2.23 $\pm$ 0.36 &  4.09 $\pm$ 0.64\\ \hline
\end{tabular}
\caption{Average adhesion energy of the different defect strip interfaces
$G_{2c}=G_0+\Delta G_2$ computed by FE analysis, perturbation theory using Rice
(semi-infinite geometry) and Legrand (plate geometry) formulas respectively.}
\label{tab:adhesion}
\end{table}

The results are summarized in Tab.~\ref{tab:adhesion}. As expected following the conclusions 
about ERR contrasts of the previous section, the semi-infinite model predictions deviate
significantly from the FE results summarized in
Tab.~\ref{tab:adhesion}. In turn, we observe a very satisfactory
agreement between the values obtained from perturbative approaches
using the plate model and the values calculated by FE. The agreement
is even quantitative, i.e. inside the error bars, for the samples
LC-b40, MC-b40 and MC-b50. Moreover, nearly identical adhesion energies are found for
the MC-b40 and MC-b50 samples due to the identical nature of their
cracked interfaces as reported in Tab.~\ref{ta:stack}. The adhesion energy for the defect
strip of sample HC-b30 is slightly out of the
margins of error calculated. This interface presents however the
largest adhesion contrast of the four samples studied. One
interpretation could therefore reside in the fact that in the latter
case we touched the limits of first-order perturbative approaches.

As a result, our approaches based on the patterning of a weak
interface can be considered as a new method to determine the
adhesion energy of heterogeneous interfaces for transparent materials by
measuring the crack front deformation that it induces.

\section{Summary and perspectives}

We have shown how the results of cleavage test experiments can be thoroughly analyzed by FEM. 
The ERR has been computed for all the measured crack front and a consistent picture of the 
crack propagation in the heterogeneous interface has been reconstructed. This analysis of the 
data is particularly illuminating when considering the increasing deformation of the crack front 
as it hits the edge of the high toughness area as it progresses forward. Due to the increasing 
curvature of the crack front, the local energy release rate rises until it reaches the 
toughness of the pinning area which then starts to rupture.

This relation between front curvature and local energy release rate is at the core of the 
perturbation method proposed by \cite{Rice1985}. Fitting the front with a perturbation kernel is a 
much lighter way of analyzing the data. Here we have demonstrated that plate thickness 
is a first order parameter in modelling crack front morphology by perturbation methods 
when the wavelengths of the perturbation are comparable to the thickness. 
In our experimental configuration, the semi-infinite kernel \citep{Rice1985} gives results which 
are only qualitatively correct, while the agreement is clearly improved when the finite 
plate thickness kernel is used \citep{Legrand2011}.

Our study, combining experiments, numerical simulations and theoretical analysis 
shows that it is possible to describe quantitatively the local adhesion
contrasts only from the observed crack front
morphologies taking into account the whole problem geometry. As a
result, it can serve as a method to characterize the local toughness
of transparent materials.

To improve the description of the depinning threshold for crack
propagation in heterogeneous brittle materials two directions of
investigation can be naturally considered from this work. The first
one is to develop an approach capable of handling the anticlastic
effect analytically \citep{Jumel2008}. Under this condition, our
approach could be more effective and avoid any numerical simulations. A
second perspective is related to the size of the perturbation
induced by the heterogeneity of the material. Elastic line models
presented in this paper are indeed limited by a first order
perturbative approach. If it seems that it can account for adhesion
contrasts up to $4$~J/m$^2$, we expect a loss of validity as the
toughness contrast increases. Approaches including higher orders in
perturbation \citep{Rice1989,Bower1991,Lazarus2003,Favier2006,Leblond2012}
could be a solution and should allow to treat much higher adhesion
contrasts.

\section*{Acknowledgements}

The support of the ANR Programme SYSCOMM (ANR-09-SYSC-006) is
gratefully acknowledged.

\bibliographystyle{model2-names}
\bibliography{Biblio}

\end{document}